\documentclass[superscriptaddress,aps,prd,nofootinbib]{revtex4-2}
\usepackage{graphicx}
\usepackage{dcolumn}
\usepackage{hyperref}
\usepackage{ulem}
\usepackage{color}
\usepackage{bm}
\usepackage{siunitx}
\usepackage{braket}
\usepackage{amsfonts,amsmath,amssymb,bm,tensor}
\usepackage{ifpdf}
\usepackage{slashed}
\usepackage[mathscr]{eucal}
\usepackage[utf8]{inputenc}
\usepackage{cancel}

\newcommand{\df}{\mathrm{d}}

\begin{document}

\begin{flushright}
TU-1135
\end{flushright}

\title{
Constraints on small-scale primordial density fluctuation
 \\
 from cosmic microwave background
 through dark matter annihilation
}

\author{Masahiro Kawasaki}
\affiliation{ICRR, University of Tokyo, Kashiwa, 277-8582, Japan}
\affiliation{Kavli IPMU (WPI), UTIAS, University of Tokyo, Kashiwa, 277-8583, Japan}
\author{Hiromasa Nakatsuka}
\affiliation{ICRR, University of Tokyo, Kashiwa, 277-8582, Japan}
\author{Kazunori Nakayama}
\affiliation{Department of Physics, Tohoku University, Sendai, Miyagi 980-8578, Japan}
\affiliation{Department of Physics, Faculty of Science, The University of Tokyo, Bunkyo-ku, Tokyo 113-0033, Japan}

\begin{abstract}

The cosmic microwave background (CMB) observation by the Planck satellite precisely determines primordial curvature fluctuations on larger scales than $\mathcal O(1)\,\mathrm{Mpc}$, while the small-scale curvature fluctuation is still less constrained.
The constraint on small-scale fluctuations is highly improved if we assume the standard thermal relic dark matter scenario. 
When small-scale fluctuations are large enough, dense regions collapse to form small halos even in a redshift $z\gtrsim10^3$, which is called ``ultracompact minihalos''.
These minihalos enhance the annihilation of the dark matter and it is constrained by observations such as extragalactic gamma rays and the CMB.
We revisit the effect of minihalos formed by the small-scale density fluctuations and calculate the ionization history modified by the dark matter annihilation.
We perform the Markov Chain Monte Carlo method to constrain the size of small-scale curvature fluctuations by the CMB power spectrum.
It is found that the constraint from the CMB power spectrum is comparable to that from the extragalactic gamma rays.
We confirm that our constraint mainly comes from the energy injection in early time ($z\gtrsim 100$) and hence it is independent of the uncertainty of minihalo properties in the late time.

\end{abstract}

\maketitle
\tableofcontents


\section{Introduction}
\label{sec_intro}

The Planck satellite precisely measured the cosmic microwave background (CMB) anisotropy and determined primordial curvature fluctuations on larger scales than $\mathcal O(1)\mathrm{Mpc}$, which shows the nearly flat power spectrum consistent with the slow-roll inflation.
On small scales, on the other hand, the sensitivity is limited due to the free streaming of photons (Silk damping), and little is known about the power spectrum.
However, the small-scale fluctuations are expected to provide hints to new physics beyond the slow-roll inflation.
Some measurements have already put upper limits on the small-scale power spectrum, e.g. 
the clustering of galaxies~\cite{BOSS:2016hvq}
the Lyman-alpha forest~\cite{SDSS:2004kjl}
the gravitational waves~\cite{Assadullahi:2009jc,Inomata:2018epa}, 
the CMB spectral distortion~\cite{Chluba:2012we}, 
and the primordial black holes~\cite{Josan:2009qn}.
These methods, however, only cover limited scales or put much loose upper limits compared to the CMB anisotropy.

The constraints on the small-scale density fluctuation are highly improved when the annihilation of dark matter particles is taken into account.
Assuming the standard thermal freeze-out scenario of the dark matter, the annihilation cross section is estimated as $\braket{\sigma v} =  3\times 10^{-26}\mathrm{cm}^3/\mathrm{s}$ for the $s$-wave annihilation.
When small-scale fluctuations are large enough, dense regions collapse to form small halos in the early universe, even in a redshift $z\gtrsim10^3$, which is often called  ``ultracompact minihalos''~\cite{Ricotti:2009bs}.
Such minihalos enhance the annihilation of the dark matter, which is severely constrained by the extragalactic gamma rays~\cite{Lacki:2010zf,Josan:2010vn,Bringmann:2011ut}.
The constraint on the small-scale power spectrum is obtained as $\mathcal P_\zeta \lesssim 10^{-6}$ over the broad scales $k\in [10^2,10^6]~\mathrm{Mpc}^{-1}$.
This constraint, however, has uncertainty related to the property of the dark matter, the formation of minihalos, and the stability of minihalos on the late time.

The CMB anisotropy measurement also constrains the dark matter annihilation signal.
The dark matter annihilation injects the energetic decay products into the thermal plasma, which ionizes the neutral hydrogen and helium.
The CMB power spectrum is sensitive to modification of the ionization history around $z\sim 600$~\cite{finkbeiner2012searching}, and it provides a reliable constraint on the exotic energy injection in the early universe.
Such constraints have been intensively investigated for the case of (almost) spatially homogeneous dark matter distribution~\cite{Padmanabhan:2005es,Mapelli:2006ej,Zhang:2006fr,Kanzaki:2008qb,Kanzaki:2009hf,Slatyer:2009yq,Slatyer:2012yq,Galli:2013dna,Slatyer:2015jla,Slatyer:2015kla,Liu:2016cnk,Acharya:2019uba,Cang:2020exa,Kawasaki:2021etm}, and the Planck satellite puts the upper bound on the annihilation cross section as $\braket{\sigma v}/m_\chi \lesssim  10^{-27}\mathrm{cm}^3/\mathrm{s}/\mathrm{GeV}$ for $s$-wave annihilation with $\chi\chi\to e^+e^-$ channel~\cite{Slatyer:2015jla}, where $\chi$ denotes the dark matter particle and $m_\chi$ is the dark matter mass.
The annihilation signal is enhanced if the dark matter distribution is (strongly) inhomogeneous, but it is known that the enhancement is negligible for the large halos formed under the nearly scale invariant primordial power spectrum predicted by the standard slow-roll inflation~\cite{Lopez-Honorez:2013cua}, which we call the ``standard halos'', since the standard halos are formed in late time.
On the other hand, large fluctuations on small scales can enhance the production of minihalos and the effect on the CMB anisotropy can also be significant.
The pioneering work~\cite{Zhang:2010cj} investigated the modified ionization history by the dark matter annihilation from ultracompact minihalos.
However, the author mainly used the profile of minihalos unfavored from the view point of the recent numerical calculation~\cite{Gosenca:2017ybi,Delos:2018ueo} and derived the constraints just by comparing the optical depth measured by the WMAP, and did not put the upper bound on the primordial curvature power spectrum.
Other works~\cite{Bringmann:2011ut,yang2011constraints,Emami:2017fiy} also inherit some incomplete treatments of Ref.~\cite{Zhang:2010cj}.
Therefore, it is necessary to consistently treat the CMB constraint on the small-scale fluctuations based on the recent updates.

In this work, we revisit the minihalo formation due to the small-scale density fluctuations and calculate the modified CMB power spectrum including the effect of dark matter annihilation.
In Sec.~\ref{sec_halo}, we describe the properties  of minihalos. Both the density profile of each minihalo and the abundance (or mass distribution) of minihalos are uncertain and hence we adopt two methods:
(i) the Navarro-Frenk-White (NFW) profile with the Press-Schechter formalism used in Ref.~\cite{Nakama:2017qac}, 
and (ii) the Moore profile with the peak theory motivated by the numerical calculation of minihalos~\cite{Delos:2018ueo}.
We compare both methods and estimate the enhancement of the dark matter annihilation, which is dubbed as the ``boost factor''.
In Sec.~\ref{sec_cmb}, the energy injection from the dark matter annihilation in the minihalos is calculated assuming the standard thermal freezeout scenario of dark matter. 
The modified ionization history of the universe is calculated by the \textsf{RECFAST}~\cite{Seager:1999km,Seager:1999bc,Wong:2007ym} code with the exotic energy injection.
We use the \textsf{CAMB}~\cite{Lewis:1999bs} and \textsf{CosmoMC}~\cite{Lewis:2002ah} code to calculate the CMB power spectrum and perform the Markov Chain Monte Carlo methods (MCMC) to put the exclusion limit of the parameters, including the small-scale curvature fluctuation.
We conclude this paper in Sec.~\ref{sec_conclusion}.

\section{Abundance and property of halos}
\label{sec_halo}

In this section, we summarize the abundance, profile, and mean squared density of minihalos.
Halos are formed when high-density regions collapse gravitationally.
The abundance and profile of the halos are estimated based on the density power spectrum, while the estimation has some uncertainty depending on methods.
We discuss two methods to estimate halos and compare their effects on the enhancement of the dark matter annihilation signal.

At first, let us introduce the small-scale density fluctuation.
Since the curvature fluctuations are constant on the superhorizon scale, it is useful to characterize the small-scale density fluctuation in terms of the curvature fluctuation.
In this paper, we use the delta-function-type power spectrum of the small-scale curvature perturbation defined by
\begin{align}
	\mathcal P_{\mathcal R} (k)
	=
	 A^2 k_*\delta(k-k_*)
	,
	\label{eq_powerspec_formula}
\end{align}
where $k_*$ is a wave number of the fluctuation and $A^2$ is an amplitude, which is assumed to be much larger than the naive value extrapolated from the flat power spectrum at the CMB scales.
We neglect the flat power spectrum to estimate the halo abundance since structure formation from the flat power spectrum is too late to affect the CMB constraints on the dark matter annihilation~\cite{Kanzaki:2009hf,Huetsi:2009ex}.
Our method is easily applied to other spiky power spectra. 
The density fluctuations are related to the curvature fluctuations as
\begin{align}
	\mathcal P_\delta (k,a)
	= \frac{4}{25} 
	\left(
		D(a)
	\frac{k^2T(k)}{\Omega_{m0} H_0^2}
	\right)^2
	\mathcal P_{\mathcal R} (k),
\end{align}
where $\Omega_{m0}\simeq 0.311$ is the current density parameter of the total matter, $D(a)$ is a growth factor as a function of the cosmic scale factor $a$ and $T(k)$ is the transfer function.
We follow Ref.~\cite{Nakama:2017qac} for a choice of them.
The growth factor is given by~\cite{Buenobelloso:2011sja}
\begin{align}
	D(a) &= a\cdot {}_2F_1
	\left[
	\frac{w-1}{2w},
	\frac{-1}{3w},
	1-\frac{5}{6w};
	1-X(a)
	\right]
	,
\end{align}
where $w$ is the equation of state parameter of dark energy (we will take $w=-1$ in what follows), ${}_2F_1$ is the Hypergeometric function, and $X(a)\equiv 1+a^3 \Omega_\Lambda/\Omega_{m0}$ with $\Omega_\Lambda$ being the density parameter of dark energy.
The transfer function without baryonic effects is given by~\cite{Nakama:2017qac, Eisenstein:1997ik}
\begin{align}
	T(q) &=
	\frac{\ln(2e+1.8q)}{\ln(2e+1.8q) +(14.2+\frac{731}{1+62.5q})q^2  }
	\quad,\quad
	q\equiv  \frac{k}{\Omega_{m0} h^2 \mathrm{Mpc}^{-1}},
\end{align}
where $h$ is the present Hubble parameter in units of $100\,{\rm km/s/Mpc}$.
The above estimation ignores the kinetic decoupling and the free streaming of the dark matter, which may washout density fluctuations on scales smaller than the cut-off scale $k_c$.
For the typical weakly interacting massive particle (WIMP) dark matter, the cut off scale is given by $k_c \sim 10^{6-7}~\mathrm{Mpc}^{-1}$~\cite{Green:2005fa,Loeb:2005pm}.
In this paper, we consider the spiky power spectrum with peak scale $k_*\in [1 ~\mathrm{Mpc}^{-1}, 10^6 ~\mathrm{Mpc}^{-1}]$.

In order to evaluate the energy flux from dark matter annihilation, we need information about the number density and the density profile of minihalos.
In the context of ultracompact minihalos, previous works often used the steep inner profile, $\rho(r)\propto r^{-9/4}$, which is analytically derived for collisionless spherical infall on to an overdense region~\cite{1985ApJS...58...39B}.
A recent N-body numerical calculation~\cite{Delos:2017thv,Gosenca:2017ybi,Delos:2018ueo}, however, found that minihalos have the shallower inner profiles with the inner slopes between $\rho(r)\propto r^{-1}$ and $r^{-3/2}$.
In this paper, we investigate the halos with both NFW profile with $\rho(r)\propto r^{-1}$ and Moore profile with $\rho(r)\propto r^{-3/2}$.

According to the numerical calculation in \cite{Delos:2018ueo}, the profile of minihalos depends on their merger history.
When we use the spiky power spectrum, the isolated minihalos are formed with the Moore profile.
The parameters of the halo profile are determined by the environment at the formation, and the parameters depend only on the formation time.  
In this case, the halo abundance is estimated by the peak theory in which the isolated peak number is counted.
On the other hand, when the power spectrum has a step-like form on small scales, the formed halos have a broad distribution on their mass and experience many times of mergers.
After mergers, the halo profile becomes shallower, and the NFW profile well fits resultant halo profiles.
In this case, the Press-Schechter formalism is suitable for counting the halo number since it is closely related to the coarse-grained density contrast and includes the effect of mergers~\cite{Bond:1990iw,Lacey:1993iv,Zentner:2006vw,Maggiore:2009rv}.

Considering the above discussion, we adopt two methods to investigate minihalos formed at high redshifts.
The first method was investigated in Ref.~\cite{Nakama:2017qac}, where the authors used the NFW profile with the Press-Schechter formalism to estimate the time-dependent concentration parameter of the halo profile, which will be defined later.
The second method was investigated in Ref.~\cite{Delos:2018ueo}, where the Moore profile was used with the peak theory to estimate the collapsed time of halos.
Although we use the delta-function type power spectrum for simplicity, the primordial power spectrum inevitably has a peak width.
The evolution of minihalos follows either of two pictures depending on the width, while a threshold between the above two pictures is not clear.
Thus, in this paper, we consider both pictures, which we simply call ``NFW profile'' and ``Moore profile''. 
The annihilation signal is evaluated on both scenarios in the following discussion.

We compare both pictures to estimate the dark matter annihilation flux.
Since the annihilation rate is proportional to the squared number density of dark matter particle, $n_\chi^2= \rho_\chi^2/m_\chi^2$, we calculate the square average of the halo density profile, ${\braket{\rho_\mathrm{halo}(z)^2}}$, 
where the bracket represents the average over halos and is defined in Eqs.~\eqref{eq_squaredaverage_NFW} and \eqref{eq_moore_braketrho2}.

\subsection{NFW profile}
\label{sec_halo_NFW}

The NFW profile is one of the standard halo profiles, which is defined as
\begin{align}
	\rho_\mathrm{NFW}(r)
	=  
	\frac{\rho_s}{(r/r_s)(1+r/r_s)^2},
	\label{eq_profile_NFW}
\end{align}
where $\rho_s$ and $r_s$ are parameters that depend on the halo mass and virial radius.
The virial radius of a halo with mass $M$ is calculated by the spherical collapse model as
\begin{align}
    r_h(z,M) =
    \left(
    \frac{3M}{4\pi \Delta_\mathrm{vir}(z) \rho_m(z)}
    \right)^{1/3},
\end{align}
where $\rho_m(z)$ is the matter density and $\Delta_\mathrm{vir}(z) $ is given by the fitting function neglecting the radiation energy density~\cite{Bryan:1997dn}:
\begin{align}
    \Delta_\mathrm{vir}(z)
    &\simeq
    \frac{18\pi^2+82x -39x^2}{1+x}
    \quad,\quad
    x\equiv
    \frac{\Omega_m(z)}{\Omega_m(z)+\Omega_\Lambda}-1
    .
\end{align}
To determine $\rho_s$, we require that the mass of the halo is given by the integration of $\rho_\mathrm{NFW}(r)$ up to the virial radius, $M = \int^{r_h} \df^3 r \rho_\mathrm{NFW}(r)$. It is given by
\begin{align}
    \rho_s  = \frac{c}{\ln(1+c) -c/(1+c)}
    \frac{\Delta_\mathrm{vir}(z) \rho_m(z)}{3},
\end{align}
with $c = r_h/r_s$ is called the concentration parameter and determined by  numerical simulations.
There are various choice of $c$~\cite{Bullock:1999he,Ullio:2002pj,Hennawi:2005bm,Comerford:2007xb}, and in this paper, we use the fitting formula by \cite{Zhao:2008wd}:
\begin{align}
    c(t) = 4\times \left[
    1+ \left(
    \frac{t}{3.75 t_{0.04}}
    \right)^{8.4}
    \right]^{1/8},
    \label{eq_concentration}
\end{align}
where $t_{0.04}$ is the time when a halo gained $4\%$ of its mass evaluated at $t$.
In summary, we can determine the halo profile by its mass $M$ and the formation time $t_{0.04}$.

Next, we evaluate the halo abundance based on the Press-Schechter formalism~\cite{Press:1973iz}.
The halos are formed by density fluctuations growing on the subhorizon regime.
Although the Sheth-Tormen mass function~\cite{Sheth:1999mn} well fits the mass spectrum of standard halos at low redshifts, it is unclear if this function is applicable to minihalos formed at high redshifts.
Thus we use the Press-Schechter formalism in this paper for simplicity.
In the Press-Schechter formalism, the formation rate of halos is estimated by the coarse-grained density fluctuation,  
\begin{align}
	\delta_R (t, \bm x) 
	= 
	\int \df^3 \bm x'
	~
	 W(|\bm x - \bm x' |,R)  
	 \delta(t, \bm x')
	 , 
\end{align}
where $W(x,R)$ is the window function.
We choose the Gaussian window function,
\begin{align}
	W(x,R) 
	&= 
	(2\pi)^{-3/2} R^{-3} \exp\left(-\frac{r^2}{2R^2} \right)
	\quad,\quad
	\tilde W(kR) 
	= 
	\exp\left(-(kR)^2/2 \right)
	,
\end{align}
where $\tilde W$ is the Fourier-transformation of $W$.
The mass and radius of the collapsed object are related as
\begin{align}
	M(R) = W(0,R)^{-1} \rho_{m0}  =  (2\pi)^{3/2} R^3\rho_{m0}  
	,
\end{align} 
with the present matter density $ \rho_{m0} $.
The halo collapses when $\delta_R $ overcomes the critical value $\delta_c=1.686$ in the linear perturbation theory.
The formation rate is estimated by assuming the Gaussian distribution of fluctuations,
\begin{align}
	\beta(M,z)  = 
	\frac{1}{\sqrt{2\pi}\sigma_M(z) }
	\int_{\delta_c}^\infty
	\df \delta 
	\exp\left(
		-\frac{\delta^2}{2\sigma^2_M(z)}
	\right)
	=
	\frac{1}{2} \mathrm{erfc}(\nu_M/\sqrt{2})
	,
	\label{eq_beta_PS}
\end{align}
where $\sigma_M(z)$ is the variance of coarse-grained density fluctuations and $\nu_M \equiv \delta_c/\sigma_M$.
The variance is estimated by
\begin{align}
	\sigma_M^2 (z)
	= \braket{\delta_R(t,\bm x)\delta_R(t,\bm x)}
	=
	 \int \frac{\df k}{k} \mathcal  P_\delta(k,z) \tilde W^2(kR).
\end{align}
Based on the Press-Schechter formalism, the mass distribution of halos is derived as 
\begin{align}
	\frac{\df n}{\df \ln M}(M,z)
	=-2\frac{\rho_{m0}}{M} \frac{\df \beta}{\df \ln M}
	=
	\sqrt{\frac{2}{\pi}}
	\frac{\rho_{m0}}{M} \frac{-\df \ln\sigma_M}{\df \ln M}
	\nu_M 
	\exp(-\nu_M^2/2),
	\label{eq_press_schechter}
\end{align}
where $n$ represents the comoving number density of halos.

\begin{figure}[th]
	\centering
	\includegraphics[width=0.55\textwidth ]{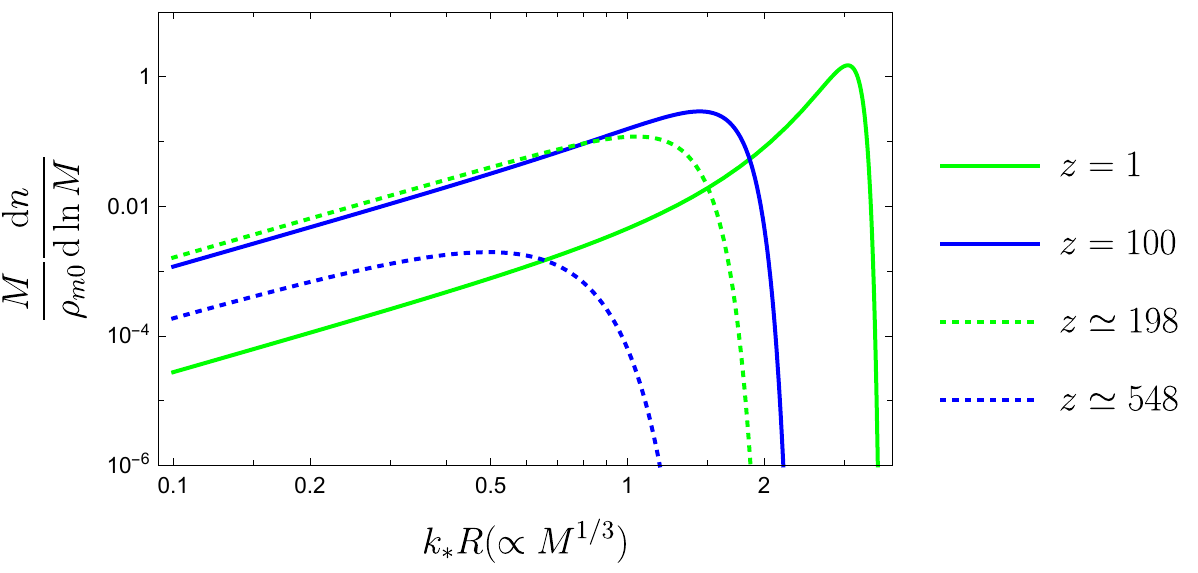}
	\includegraphics[width=0.4\textwidth ]{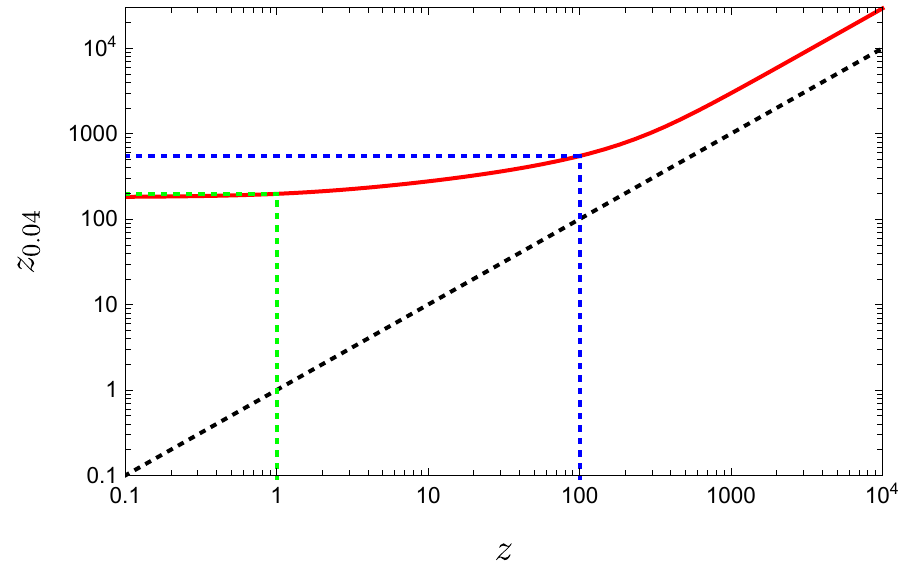}
	\caption{
	The mass distribution of minihalos calculated by the Press-Schechter formalism (left panel) and the redshift of the halo formation (right panel).
	In both case, we use the power spectrum~\eqref{eq_powerspec_formula} with $A^2 = 10^{-6}$ and $k_*=10^3~\mathrm{Mpc}^{-1}$.
	In the left panel, the dotted lines represent the mass distribution when the typical halo mass is $4\%$ times smaller than that of solid lines.
	The red line of the right panel shows the formation time $z_{0.04}$ relative to the observed redshift $z$.
	The blue and green dotted lines correspond to the redshift in the left panel. 
	The black dotted line represents $z_{0.04} = z$ as a comparison.
	}
	\label{fig_PS_mass_dist}
\end{figure}

Using the mass distribution, we estimate the formation time $t_{0.04}$ following the method in Ref.~\cite{Nakama:2017qac}.
The normalized mass distribution, $f(M,z)\equiv ({M}/{\rho_{m0}})\frac{\df n}{\df \ln M}$, has a peak at mass $M_c(z)$ defined by $[\df f(M,z)/\df M]|_{M=M_c} =0 $.
$M_c(z)$ represents the typical halo mass at redshift $z$.
We need to calculate the redshift $z_{0.04}$ when $M_c(z_{0.04})$ is 4\% of $M_c(z)$.
We define the halo formation redshift $z_{0.04}(z)$ relative to the halo at $z$ by $M_c(z_{0.04}(z)) = 0.04M_c(z)$.
We show $z_{0.04}(z)$ for $z=1$ and $100$ in Fig.~\ref{fig_PS_mass_dist} based on the power spectrum with $A^2 = 10^{-6}$ and $k_*=10^3~\mathrm{Mpc}^{-1}$.
The left panel of Fig.~\ref{fig_PS_mass_dist} shows the normalized mass distribution at $z=1$ and $100$ (solid lines) .
We also show the mass distribution at $z\simeq 198$ and $548$ (dotted lines) where their typical masses are $0.04$ times smaller than these at $z=1$ and $100$, respectively.
The right panel of the Fig.~\ref{fig_PS_mass_dist} shows the relation between $z_{0.04}(z)$ and $z$.
The formation redshift $z_{0.04}$ is much larger than $z$ on low redshift due to the large density fluctuations. 
Using this relation, we can calculate the concentration parameter~\eqref{eq_concentration}. 

We comment on the effect of baryons on the halo formation.
In the above analysis, we use the total matter density to calculate the halo abundance.
When the halo has a mass smaller than the Jeans mass, the pressure prevents the baryons from collapsing, which decreases the density contrast by about $0.05/0.3\sim 17\%$ compared to the above analysis.
In this paper, we neglect this effect and equally treat the dark matter and baryon in the halo formation.
We discuss this effect on the dark matter constraint in Sec.~\ref{sec_discussion}.

The energy flux from dark matter annihilation is proportional to the square of the dark matter density averaged over the halo profile and the physical number density of halos, which is given by
\begin{align}
    \braket{\rho_\mathrm{halo}(z)^2}
    &\equiv
    \int \df M 
    {(1+z)^3}
    \frac{\df n}{\df M}(M,z)
    \int^{r_h}\df^3 r \rho_\mathrm{halo}(r;M,z)^2,
    \label{eq_squaredaverage_NFW}
\end{align}
where $\rho_\mathrm{halo}(r;M,z)$ represents the dark matter distribution in a halo with mass $M$ at redshift $z$.
Here we simply assume that the dark matter distribution is proportional to the matter distribution in the halo, $\rho_\mathrm{halo} = (\Omega_{c0}/\Omega_{m0})\rho_\mathrm{NFW} $, with the current dark matter density parameter $\Omega_{c0}$.
The radial integration up to the virial radius is calculated as
\begin{align}
	\int^{r_h}\df^3 r
	\rho_\mathrm{halo}(r,M,z)^2
	&=
	\left( \frac{\Omega_{c0}} {\Omega_{m0}} \right)^2
	\int^{r_h}\df^3 r
	\rho_\mathrm{NFW}(r)^2
	\\
	&=
	\left( \frac{\Omega_{c0}} {\Omega_{m0}} \right)^2
	\Delta_\mathrm{vir} \rho_m 
	M f_\mathrm{NFW}(c),
	\\
	f_\mathrm{NFW}(c)
	&\equiv
	\frac{c^3}{9}
	\frac{ 1-1/(1+c^3)  }{ [\ln(1+c)-c/(1+c) ]^2 },
\end{align}
and the squared average of the dark matter density is given by 
\begin{align}
	\braket{\rho_\mathrm{halo}(z)^2}
	&=
	\int \df M~ 
	 {(1+z)^3}
	\frac{\df n}{\df M}(M,z) 
	~
	\left( \frac{\Omega_{c0}} {\Omega_{m0}} \right)^2
	\Delta_\mathrm{vir} \rho_m(z)
	M f_\mathrm{NFW}(c)
	\\&=
	2
	\left( \frac{\Omega_{c0}} {\Omega_{m0}} \right)^2
	\Delta_\mathrm{vir} \rho_m(z)^2
	f_\mathrm{NFW}(c(z))
	\beta(M=0,z).
	\label{eq_rhohalo2_NFW}
\end{align}

\subsection{Moore profile}
\label{sec_halo_Moore}

In this section, we adopt the simulation result and analysis in Ref.~\cite{Delos:2018ueo}.
The authors suggest that the profile of minihalos follows the Moore profile:
\footnote{
    Note that this profile is slightly different from the original profile suggested by Moore~et.al.~\cite{Moore:1999gc}, 
    \begin{align}
        \rho(r) = \frac{\rho_s}{(r/r_s)^{3/2} (1+(r/r_s)^{3/2})}.
        \nonumber
    \end{align}
    but we call Eq.~\eqref{eq_profile_moore} ``Moore profile'' in this paper. 
}
\begin{align}
	\rho_\mathrm{Moore}(r) \equiv \frac{\rho_s}{(r/r_s)^{3/2} (1+r/r_s)^{3/2}}.
	\label{eq_profile_moore}
\end{align}
The parameters of the profile are estimated by the numerical calculation using a spiky power spectrum with a peak scale $k_*$ as
\begin{align}
	r_s &= \left(\frac{17}{30}\right)^{2/3} k_*^{-1} a_c 
	\quad,\quad
	\rho_s = 30  a_c^{-3}
	\rho_{m0}
	,
	\label{eq_haloprof_moore_param}
\end{align}
where $a_c$ is the scale factor at the halo formation time ($a_0 =1)$.

Compared to the NFW profile in the previous subsection, the estimation of $\braket{\rho_\mathrm{halo}(z)^2}$ is different in two points.
First, we use the peak theory instead of the Press-Schechter formalism.
The simulation in Ref.~\cite{Delos:2018ueo} suggests that the spiky spectrum leads to isolated halos and their mergers are rare.
In this case, the peak theory is favored since the merger of halos is ignored in the peak theory [see Ref.~\cite{Delos:2018ueo} for more details].
Second, the Moore profile is steeper than the NFW profile, and the average of the dark matter density squared diverges at the center of the halos if we naively use Eq.~\eqref{eq_profile_moore}.
In reality, the dark matter density has some upper bound due to the effect of dark matter annihilation, which will {}be discussed later.

\begin{figure}[th]
	\centering
	\includegraphics[width=0.7\textwidth ]{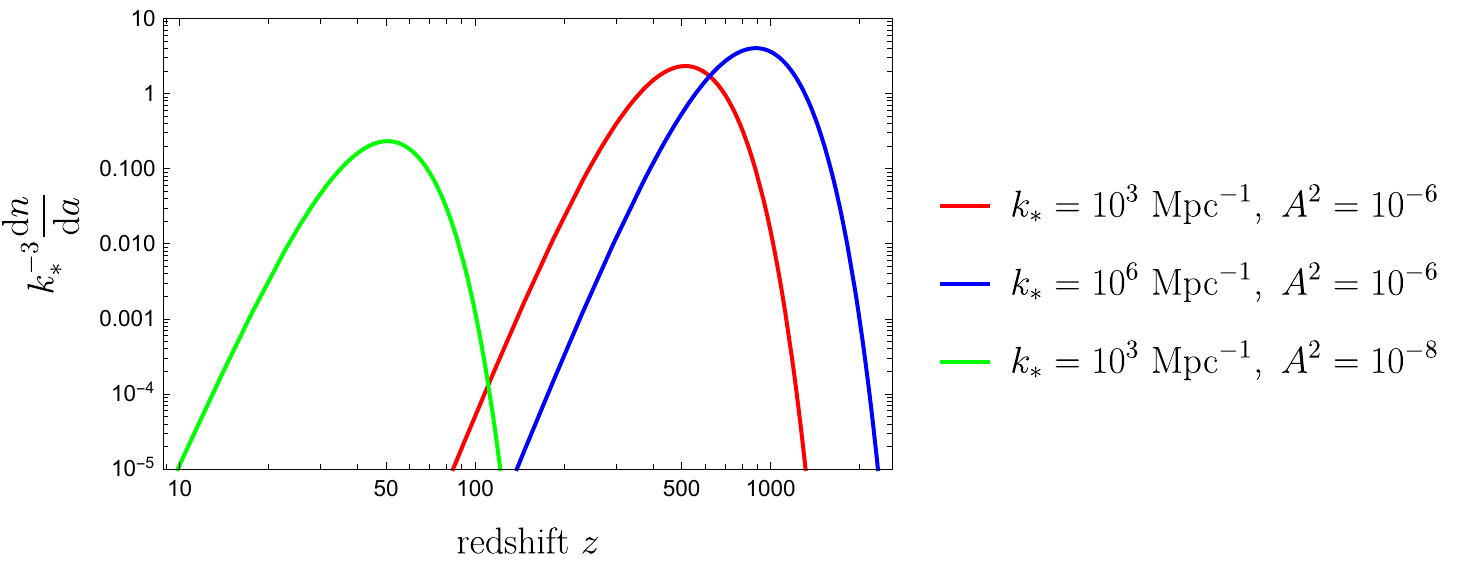}
	\caption{
	The number density of halos calculated by the peak theory in Eq.~\eqref{eq_dnda_PT}.
	The horizontal axis represents a redshift of the halo formation.
	We use three different parameters of the power spectrum.
	}
	\label{fig_Peaktheory_mass_dist}
\end{figure}

Let us calculate the halo abundance based on the peak theory investigated by Bardeen, Bond, Kaiser, and Szalay~\cite{Bardeen:1985tr} (hereafter we call it BBKS).
The peak theory calculates the probability that random Gaussian fluctuations form a peak.
A halo is formed when a height of the peak overcomes the threshold value $\delta_c=1.686$.
Using the delta-function type power spectrum, the comoving number density of halos at redshift $z$ is calculated by~\cite{Delos:2018ueo}
\begin{align}
	n(z)&=
	\frac{k_*^3}{(2\pi)^2 3^{3/2}}
	\int^{\infty}_{\delta_c /\sigma_0(z)}
	e^{-\nu^2/2} f(\nu) \df \nu,
	\label{eq_num_peak_theory}
\end{align}
where $f(\nu)$ is the function defined by Eq.~(A15) of BBKS, and $\sigma_0(z)$ is the variance of the density power spectrum,
\begin{align}
    \sigma_0(z)^2
 	&= 
 	\int \frac{\df k}{k} \mathcal  P_\delta(k,z).
\end{align}
The differential number density at the formation time is given by
\begin{align}
	\frac{\df n}{\df a_c} 
	&=
	\frac{k_s^3}{a_c}
		h\left(  \frac{\delta_c}{\sigma_0(z_c)}   \right)
		,
	\label{eq_dnda_PT}
	\\
	h(\nu)
	&=
	\frac{\nu}{(2\pi)^2 3^{3/2}}
	e^{-\nu^2/2} f(\nu) \df \nu
	,
\end{align}
where $z_c$ is the redshift of the halo formation $z_c = a_c^{-1}-1$ and we assume the linear growth of the matter fluctuation, $\sigma_0\propto a$.
We show the typical distribution of the formation redshift in Fig.~\ref{fig_Peaktheory_mass_dist}.
As $k_*$ and $A^2$ are smaller, the halo formation becomes later.
Some halos are formed even in $z>10^3$ for  $k_*=10^{6}~\mathrm{Mpc}^{-1}$ and $A^2=10^{-6}$. 

We estimate the averaged dark matter density squared using the Moore profile, which is given by
\begin{align}
    \braket{\rho_\mathrm{halo}(z)^2}
    &\equiv
    \int_a^0 \df a_c
    {(1+z)^3}
    \frac{\df n}{\df a_c}(a_c)
    \int^{r_h}_{r_\mathrm{min}}
    \df^3 r
    \rho_\mathrm{halo}(r;a_c, z)^2
    \label{eq_moore_braketrho2}
    .
\end{align}
Compared to the NFW profile case given by Eq.~\eqref{eq_squaredaverage_NFW}, the average is performed for the formation time $a_c$, and the $r$-integration has the lower bound for the following reason.
It is known that the annihilation of the dark matter is saturated in the central cusp region of the halo.
When the cusp region has a number density $n$, the typical lifetime of the dark matter in the region is $\Delta t\sim (\braket{\sigma v}n)^{-1}$.
Then, the maximum density of the core is given by~\cite{Berezinsky:1992mx}
\begin{align}
	\rho_\mathrm{max} 
	&= 
	\frac{m_\chi}{\braket{\sigma v}(t-t_i)},
\end{align}
where $t_i$ is the time of the halo formation.
We define $r_\mathrm{min}$ by $(\Omega_{c0}/\Omega_{m0})\rho_\mathrm{Moore}(r_\mathrm{min}) = \rho_\mathrm{max} $, which is given by
\begin{align}
	\epsilon 
	&\equiv   
    \frac{r_\mathrm{min}}{r_s}
\simeq 
	\left(
	\frac{\Omega_{m0}}{\Omega_{c0}}
	\frac{\rho_\mathrm{max}}{\rho_s}
	\right)^{-3/2}
	\simeq 
	6\times 10^{-11}
	a_c^{-2}
	\left(
	\frac{ 3\times 10^{-26}\mathrm{cm}^3/\mathrm{s}/\mathrm{TeV} }{\braket{\sigma v}/m_\chi}
	\frac{13.8~\mathrm{Gyr}}{t}
	\frac{30 }{ f_2  }
	\frac{
		1.27\times 10^{-6} 
		\mathrm{GeV}/\mathrm{cm}^{3}
	}{\rho_{\chi0} }
	\right)^{-2/3},
\end{align} 
and we safely take $\epsilon\ll 1$ for our calculation.

We simply assume that the dark matter distribution in halos also follows the Moore distribution as $\rho_\mathrm{halo} = (\Omega_{c0}/\Omega_{m0})\rho_\mathrm{Moore} $.\footnote{
There are ambiguities in the baryon distribution and its feedback to the dark matter distribution. 
If the halo mass is smaller the Jeans mass baryons cannot collapse into dark matter halos.
On the other hand, in large halos baryons are cooled and take different distribution, which could affect dark matter distribution. 
We neglect these ambiguities here}
The radial integration is estimated as
\begin{align}
	\int^{r_h}_{r_\mathrm{min}}
	\rho_\mathrm{halo}(r)^2
	&=
	\left( \frac{\Omega_{c0}} {\Omega_{m0}} \right)^2
	\int^{r_h}_{r_\mathrm{min}} 
	\df^3 r~
	\rho_\mathrm{Moore}(r; a_c, z)^2
	\\
	&\simeq 
	\left( \frac{\Omega_{c0}} {\Omega_{m0}} \right)^2
	2\pi \rho_s^2 r_s^3
	\left[
		2\log\left(1+\frac{1}{\epsilon}\right)
		-\frac{3+2\epsilon }{(1+\epsilon)^2}
	\right]
\\
    &\simeq 
	\left( \frac{\Omega_{c0}} {\Omega_{m0}} \right)^2
	2\pi \rho_s^2 r_s^3
	\left[
	2\log\left(\epsilon^{-1}\right)
	-3
	\right]
	,
\end{align}
where the choice of the upper bound of the integration is negligible as long as $r_h\gg r_s$ and we use $r_h\to \infty$ in this calculation.
Then, the average of the dark matter density squared is given by
\begin{align}
	\braket{\rho_\mathrm{halo}(z)^2}
	&=
	2\pi \left( \frac{\Omega_{c0}} {\Omega_{m0}} \right)^2
	\int_0^a \df a_c~ 
	\left(
	k_s^{-3}\frac{\df n}{\df a_c}(a_c)
	\right)
	~
	\rho_s^2 ( r_s k_s /{a})^3
	\left[
	2\log\left(\epsilon^{-1}\right)
	-3
	\right].
	\label{eq_rhohalo2_Moore}
\end{align}

\subsection{Comparison}
\label{sec_halo_Comparison}

Let us compare two pictures of halo formation: the NFW profile with the Press-Schechter formalism in Sec.~\ref{sec_halo_NFW} and the Moore profile with the peak theory in Sec.~\ref{sec_halo_Moore}.
The Moore profile has the denser core than the NFW profile, which enhances the annihilation flux from minihalos.
The total annihilation flux also depends on the abundance of minihalos and it is different between the Press-Schechter formalism and the peak theory.
We quantitatively estimate these points to specify the differences between the two pictures.

At first, we compare the annihilation flux from each minihalo for the NFW and Moore profiles.
Fixing the halo parameter $r_s$ and $\rho_s$, the ratio of the density squared for the Moore profile to that for the NFW profile is given by 
\begin{align}
    \frac{\int^{\infty}_{r_\mathrm{min}}
	\rho_\mathrm{Moore}(r)^2  
	}{\int^{\infty}_{0}
	\rho_\mathrm{NFW}(r)^2
	}
	=
	\frac{
		2\pi \rho_s^2 r_s^3
	\left[
	2\log\left(\epsilon^{-1}\right)
	-3
	\right]
	}{\frac{4\pi}{3} \rho_s^2 r_s^3}
	=
	\frac{3}{2}
	\left[
	2\log\left(\epsilon^{-1}\right)
	-3
	\right]
	\sim
	10^2
	,
\end{align}
where we estimate $\epsilon \equiv r_\mathrm{min}/r_s$ with $t=t(z=100)$, $a_c=10^3$ and $\braket{\sigma v}/m_\chi = 3\times 10^{-26}\mathrm{cm}^3/\mathrm{s}/\mathrm{TeV}$.
Thus, as for the annihilation flux from a single halo, the Moore profile leads to the signal about $10^2$ times larger than the NFW profile due to its dense core.

\begin{figure}[th]
	\centering
	\includegraphics[width=0.8\textwidth ]{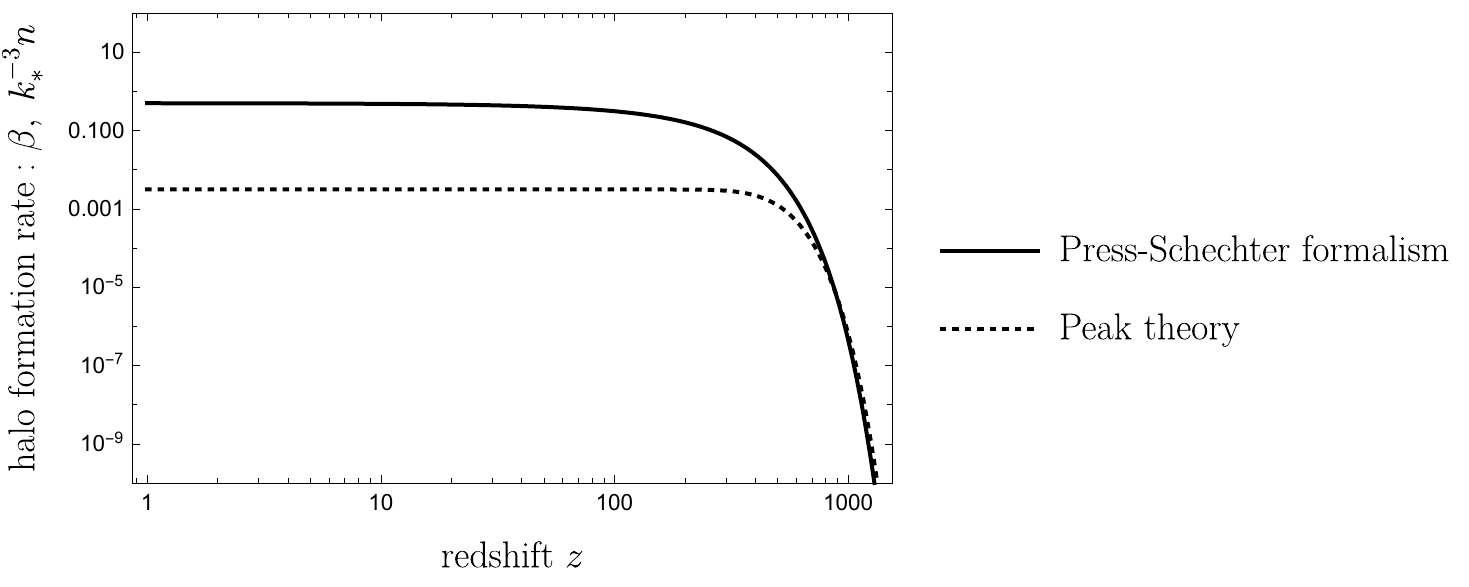}
	\caption{
	The halo formation rate for the Press-Schechter formalism and the peak theory.
	The solid line represents $\beta(M=0,z)$ in Eq~\eqref{eq_beta_PS} calculated by the Press-Schechter formalism.
	The dotted line represents a normalized number density calculated by the peak theory, $k_*^{-3}n(a)$ in Eq.~\eqref{eq_num_peak_theory}.
	}
	\label{fig_FormationRate_Compare}
\end{figure}

Second, we compare the formation rate of minihalos in the Press-Schechter formalism and the peak theory.
From Eq.~\eqref{eq_rhohalo2_NFW}, the annihilation signal is proportional to $\beta(M=0,z)$ in the Press-Schechter formalism.
We regard $\beta(M=0,z)$ as the typical halo formation rate since it is written by the mass-weighted number density per volume,
\begin{align}
    \int_0^\infty \frac{\df M}{M}  ~
    \left(
    \frac{M}{2\rho_{m0}}
    \right)
    \frac{\df n}{\df \ln M}(M,z)
    =
    \left[-\beta(M,z)\right]_0^\infty
    =
    \beta(M=0,z)
    .
\end{align}
On the other hand, the formation rate in the peak theory is regarded as $k_*^{-3}n(a)$ from Eq.~\eqref{eq_num_peak_theory}.
We show two formation rates in Fig.~\ref{fig_FormationRate_Compare}, which is evaluated for the power spectrum with $A^2 = 10^{-6}$ and $k_*= 10^3~\mathrm{Mpc}^{-1}$.
The Press-Schechter formalism predicts the formation rate about $10^2$ larger than that of the peak theory at $z \lesssim 300$. It reflects the different pictures of the halo formation.
In the peak theory, the halos are mainly formed with typical scale $k=k_*$ and then stay separated.
On the other hand, the Press-Schechter formalism includes the effects of the merger and the growth of halos, which leads to the larger halo formation rate in later times.

\begin{figure}[th]
	\centering
	\includegraphics[width=0.8\textwidth ]{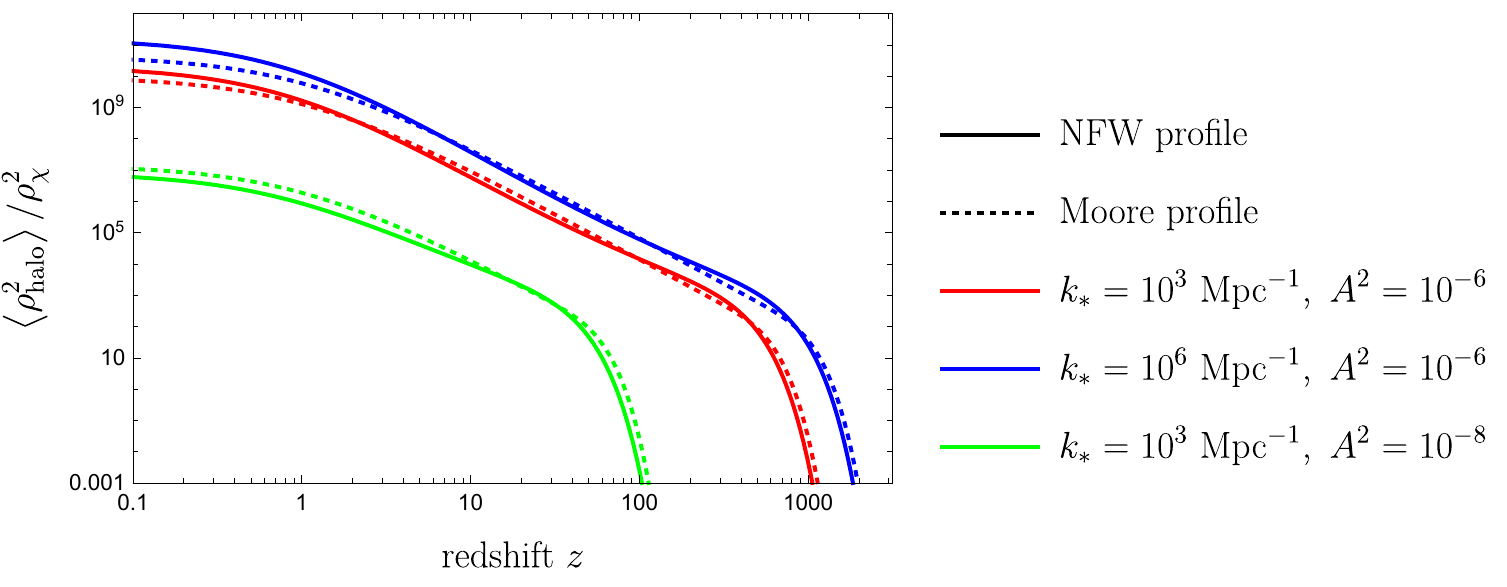}
	\caption{
	The squared average of the dark matter energy density normalized by the homogeneous dark matter density.
	The solid lines represent the results of the NFW profile with the Press-Schechter formalism~\eqref{eq_rhohalo2_NFW}, and dotted lines represent the Moore profile with the peak theory~\eqref{eq_rhohalo2_Moore}, respectively.
	The colors represent the different parametrization of the power spectrum.
	}
	\label{fig_compare_boostfactor}
\end{figure}

In summary, the NFW profile with the Press-Schechter formalism predicts a larger number of halos with smaller dark matter density than those of the Moore profile with the peak theory.
The two differences have opposite effects on the total annihilation flux.
We clarify this point by comparing the average of the dark matter density squared in Fig.~\ref{fig_compare_boostfactor}.
The solid lines represent the results of the NFW profile with the Press-Schechter formalism~\eqref{eq_rhohalo2_NFW}, and dotted lines represent the Moore profile with the peak theory~\eqref{eq_rhohalo2_Moore}, respectively.
We show the results with three parameters of the power spectrum to check the dependence on the scale and amplitude (red, blue, green lines).
In both pictures, the halo formation occurs faster when the amplitude $A^2$ is larger and the peak scale $(k_*)^{-1}$ is smaller.
Interestingly, both pictures predict similar annihilation fluxes.
Therefore, we expect that these two pictures result in a similar constraint on the small-scale curvature fluctuation. We will confirm it in the next section.

\section{Modified thermal history due to dark matter annihilation}
\label{sec_cmb}

So far we have discussed the dark matter halo profile and abundance under the small-scale density fluctuation. 
Now let us see the effect of dark matter annihilation on the thermal history of the universe taking account of the halo profile and abundance.
The injected energy due to dark matter annihilation per physical time and volume consists of two parts,
\begin{align}
	\left[ 
	\frac{\df E}{\df V\df t}
	\right]
	=
	\frac{\df E}{\df V\df t}\bigg|_\mathrm{BG}
	+
	\frac{\df E}{\df V\df t}\bigg|_\mathrm{halo}
	,
\end{align}
where the former comes from the spatially homogeneous dark matter component and the latter comes from the inhomogeneous component existing in minihalos.
The homogeneous part is given by
\begin{align}
	\frac{\df E}{\df V\df t}\bigg|_\mathrm{BG}
	=
	m_\chi (n_\chi (z))^2 \braket{\sigma v}
	=
	\frac{(\rho_\chi(z))^2}{m_\chi} \braket{\sigma v}
	,
	\label{eq_injectenergy_BG}
\end{align}
with the number density of the homogeneous component of dark matter $n_\chi (z)$. 
The energy injection from the dark matter halos is given by 
\begin{align}
	\frac{\df E}{\df V\df t}\bigg|_\mathrm{halo}
	&=
	\int \df \alpha~ 
	\frac{\df n}{\df \alpha}(\alpha,z)
	F_1(\alpha,z),
	\\
	F_1(\alpha,z)
	&=
	m_\chi \braket{\sigma v}
	\int^{R}\df^3 r
	\left(\frac{\rho_\mathrm{halo}(r,\alpha,z)}{m_\chi}\right)^2,
\end{align}
where ${\df n}/{\df \alpha}$ is the number density of halos differentiated with respect to the halo parameter $\alpha$ and $F_1(\alpha,z)$ is the annihilation energy flux from halos.
The halo parameter $\alpha$ is the halo mass for the NFW profile and the formation time for the Moore profile used in Sec.~\ref{sec_halo}.

We define the boost factor of the annihilation energy flux as
\begin{align}
	B(z) 
	&=
	\frac{\df E}{\df V\df t}\bigg|_\mathrm{halo}
	\bigg /
	\frac{\df E}{\df V\df t}\bigg|_\mathrm{BG}
	\\&=
	\int \df M~ 
	\frac{\df n}{\df \alpha}(\alpha,z)
	\int^{R}\df^3 r
	\left(\frac{\rho_\mathrm{halo}(r,\alpha,z)}{\rho_\chi(z)}\right)^2
	\\&=
	\frac{\braket{\rho_\mathrm{halo}(z)^2}}{\rho_\chi(z)^2}
    .
    \label{eq_boost_factor}
\end{align}
The boost factor is shown in Fig.~\ref{fig_compare_boostfactor} for the NFW profile~\eqref{eq_rhohalo2_NFW} and the Moore profile~\eqref{eq_rhohalo2_Moore}.
The inhomogeneous part dominates the energy injection even when the redshift is larger than $100$ if the small-scale perturbations are large enough.
Using this boost factor, we can estimate the effect of minihalos on the energy injection into the thermal plasma.

\begin{figure}[t]
	\centering
	\includegraphics[width=0.8\textwidth ]{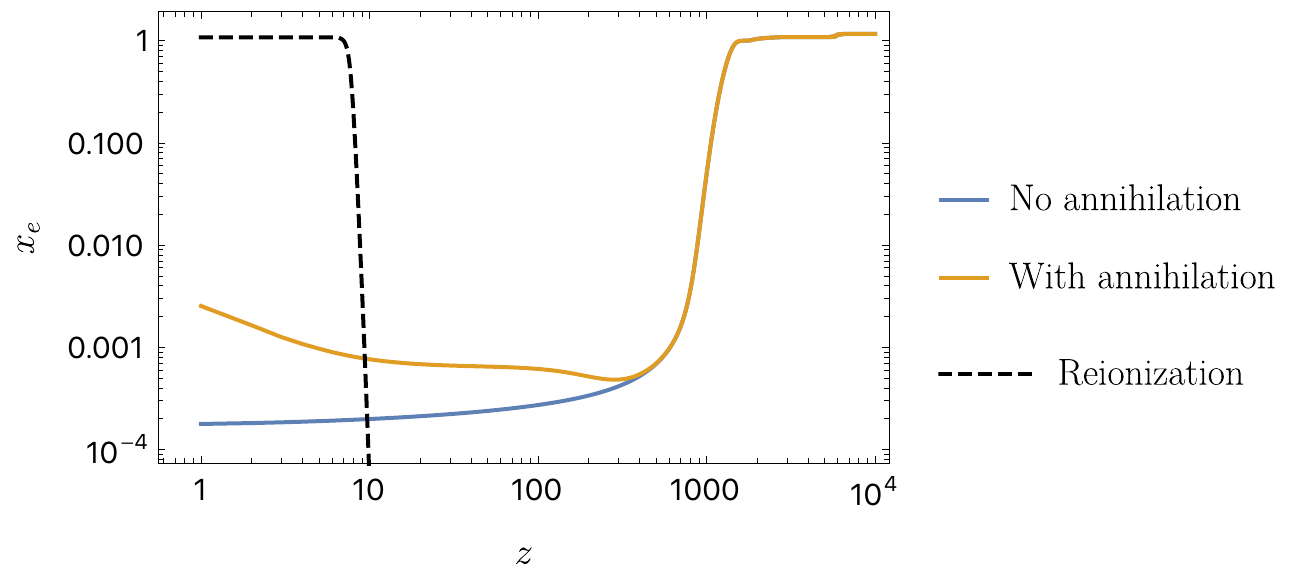}
	\caption{
	The ionization history of the universe with and without the energy injection by the annihilation of the inhomogeneous dark matter.
	$x_e$ is a ratio of the free electron number density to the hydrogen number density.
	The orange line is calculated including the dark matter annihilation with $m_\chi=1~\mathrm{TeV}$, $\braket{\sigma v} =  3\times 10^{-26}\mathrm{cm}^3/\mathrm{s}$, the $\chi\chi\to e^+e^-$ channel decay, and 
	the inhomogeneous dark matter distribution with the same boost factor to the red solid line in Fig.~\ref{fig_compare_boostfactor}.
	The black dotted line represents reionization of the universe with the \textsf{tanh} model used in Planck 2018~\cite{Lewis:2008wr, Planck:2018vyg} with $z_\mathrm{re} \simeq 7.7$.  
	}
	\label{fig_thermal}
\end{figure}

The injected particles experience various scattering processes in the background thermal plasma and lose their energy.
A part of the annihilation energy is finally used for heating of the thermal gas and ionizing hydrogen and helium.
Roughly speaking, about one-third of the injected energy is used for the ionization while its ratio depends on the energy of the injected particle, species, and the injected redshift.
We define the consumption ratio of the injected photon or electron/positron energy as $Q^{(e,\gamma)}_\alpha (E,z_\mathrm{in},z_\mathrm{abs})/E$, where $E$ is kinetic energy of the injected particle, $z_\mathrm{in}$ is the injected redshift, $z_\mathrm{abs}$ is the redshift when the energy is used, 
and $Q^{(e,\gamma)}_\alpha (E,z_\mathrm{in},z_\mathrm{abs})$ is a consumed energy at $z_\mathrm{abs}$.
The subscript $\alpha$ represents the consumption channels, $\alpha=$heat, ion${}_\mathrm{H}$, and ion${}_\mathrm{He}$. 
We use $Q^{(e,\gamma)}_\alpha (E,z_\mathrm{in},z_\mathrm{abs})/E$ calculated in our previous work~\cite{Kawasaki:2021etm}, which includes various scattering processes and is consistent with other work~\cite{Slatyer:2015kla}.
When the dark matter has various annihilation channels, we average the injected energy by their branching ratio,
\begin{align}
    Q_\alpha (z,z')
    &=
    \int \df E' 
    \bigg [
    \left( \frac{\df N_F^{(e^+)} }{\df E'} +\frac{\df N_F^{(e^-)} }{\df E'}  \right)
    Q^{(e)}_\alpha (E',z,z')
    +
    \frac{\df N_F^{(\gamma)}  }{\df E'}
    Q^{(\gamma)}_\alpha (E',z,z')
    \bigg ],
\end{align}
where  ${\df N_F^{(a)}(E') }/{\df E'}$ is the differential number density of the final state particles with the energy $E'$ for  $a=e^-,~e^+,$ and $\gamma$.
In this paper, we only consider the $\chi\chi\to e^+e^-$ channel for simplicity.
Then, the dark matter annihilation modifies the evolution of the ionization fraction of hydrogen $x_\text{H}$, that of helium $x_\text{He}$ and the gas temperature $T_M$ as
\begin{align}
	-\left[ \frac{\df x_\mathrm{H}}{\df z} \right]_\mathrm{DM}
	&=
	\int_z \frac{\df z'}{H(z')(1+z')}
	\frac{1}{n_\mathrm{H}(z')B_\mathrm{H}}
	\left[\frac{\df E(z') }{\df V\df t}\right]
	\frac{
		Q_{\mathrm{ion}_\mathrm{H}}  (z',z)
	}{m_\chi}
	,
	\label{eq_xH}
	\\
	-\left[ \frac{\df x_\mathrm{He}}{\df z} \right]_\mathrm{DM}
	&=
	\int_z
	\frac{\df z'}{H(z')(1+z')}
	\frac{1}{ n_\mathrm{He}(z') B_\mathrm{He}}
	\left[\frac{\df E(z') }{\df V\df t}\right]
	\frac{
		Q_{\mathrm{ion}_\mathrm{He}}  (z',z)
	}{m_\chi}
	,
	\label{eq_xHe}
	\\
	-\left[  \frac{\df T_M}{\df z}  \right]	_\mathrm{DM}
	&=
	\int_z
	\frac{\df z'}{H(z')(1+z')}
	\frac{2}{3}
	\frac{1}{ (1+f_\mathrm{He}+x_e) n_{H}(z') }
	\left[\frac{\df E(z') }{\df V\df t}\right]
	\frac{Q_{\mathrm{heat}}  (z',z)}{m_\chi}.
	\label{eq_heat}
\end{align}
We include these modifications in the \textsf{RECFAST} code~\cite{Seager:1999km,Seager:1999bc,Wong:2007ym} and calculate the thermal history of the universe.
The modified ionization history is shown in Fig.~\ref{fig_thermal}, where we treat the energy injection as a perturbation and use the cosmological parameters based on the Planck 2018 results~\cite{Planck:2018vyg}.
The vertical axis, $x_e$, is a number fraction of the free electron to the hydrogen atom.
The ionization ratio decreases at $z\sim 10^3$ due to the recombination of the hydrogen.
Without the energy injection from dark matter annihilation, the ionization fraction monotonically decreases (blue line) until the reionization epoch at $z_\mathrm{re}\sim 7$.
When the dark matter annihilation is taken into account, the injected energy ionizes the hydrogen and helium, and the ionization fraction increases.
The orange line shows the result for $\braket{\sigma v} =  3\times 10^{-26}~\mathrm{cm}^3/\mathrm{s}$, $m_\chi= 1~\mathrm{TeV}$, and the same boost factor as the red solid line in Fig.~\ref{fig_compare_boostfactor}.

We comment on the reionization of the universe.
We use the \textsf{tanh} model of reionization used in Planck 2018~\cite{Lewis:2008wr, Planck:2018vyg} where the redshift of the reionization $z_\mathrm{re}$ is a free parameter.
The Planck 2018 estimates $z_\mathrm{re}\simeq 7.7$ ~\cite{Planck:2018vyg} without the exotic energy injection.
One might mind whether there is a parameter degeneracy between $z_\mathrm{re}$ and the dark matter annihilation rate.
We find that the degeneracy is solved by the scale dependence of the CMB power spectrum, which will be discussed in the next section.
The reionization is also investigated by observations of high redshift objects like quasars, which suggest that the neutral fraction of the universe is $\mathcal O(10)\%$ around $z\sim 7.5$~\cite{Banados:2017unc,Davies:2018yfp,Mason:2017eqr}.
The energy injection modifies the thermal history by less than $\mathcal O(1)\%$, and thus our calculations are consistent with the observations of the neutral hydrogen.

\begin{figure}[th]
	\centering
	\includegraphics[width=0.8\textwidth ]{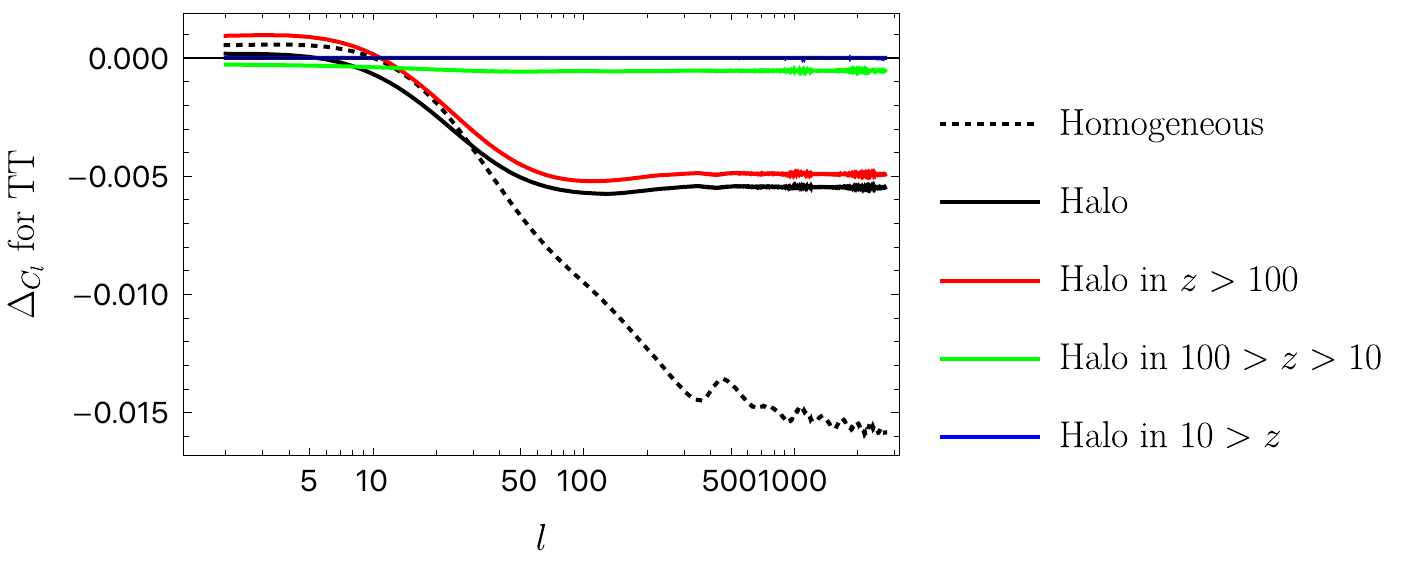}
	\includegraphics[width=0.8\textwidth ]{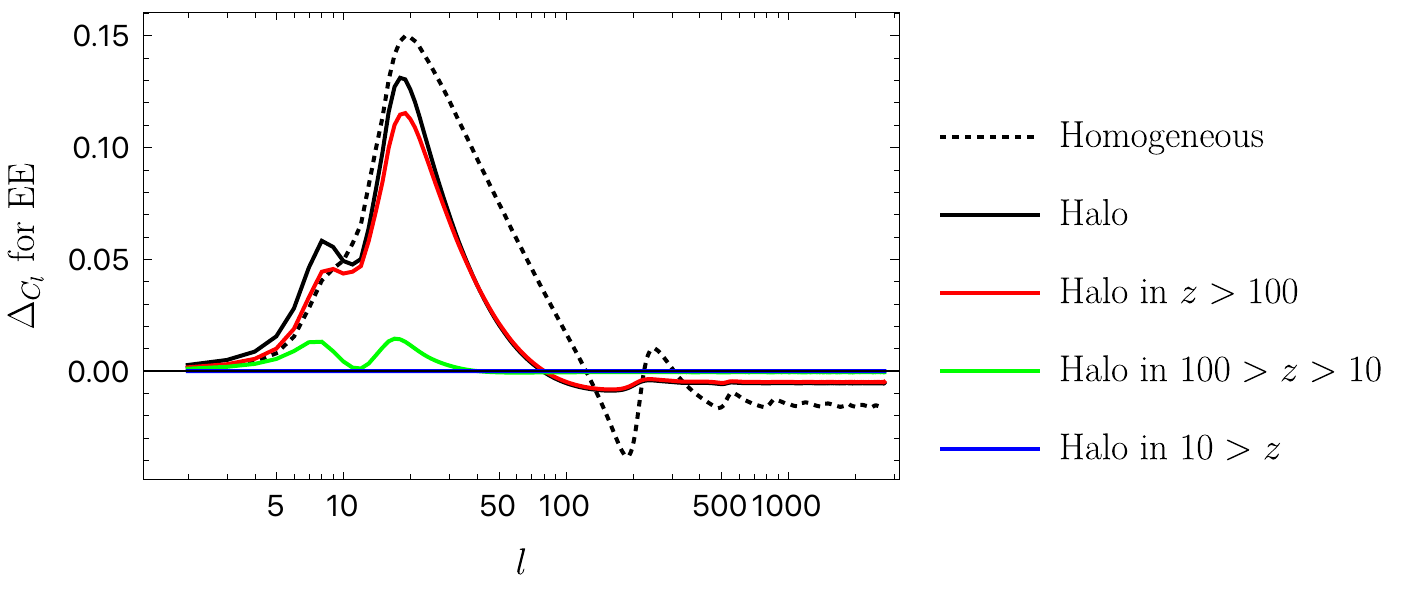}
	\caption{
	The modification of the CMB power spectrum induced by the energy injection from the annihilation of the dark matter with $m_\chi=1~\mathrm{TeV}$ and the $\chi\chi\to e^+e^-$ channel decay.
	The top and bottom panels represent the power spectrum of the temperature and the $E$-mode polarization.
	The energy injection is calculated for the homogeneous dark matter with $\braket{\sigma v} =  1\times 10^{-24}\mathrm{cm}^3/\mathrm{s}$ (black dotted line) and for the inhomogeneous dark matter with 	$\braket{\sigma v} = 3\times 10^{-26}\mathrm{cm}^3/\mathrm{s}$ and the same boost factor to the red solid line in Fig.~\ref{fig_compare_boostfactor}.
	We also show the power spectrum by artificially limiting the energy consumption epoch: early time ($z>100$, red line), middle time ($100>z>10$, green line), and late time ($10>z$, blue line), respectively.
	}
	\label{fig_delCl}
\end{figure}

The modified recombination history as discussed above affects the power spectrum of the CMB anisotropy.
The CMB photons are scattered by free electrons through the Thomson scattering process. 
The optical depth due to the Thomson scattering is given by~\cite{Planck:2018vyg}
\begin{align}
    \tau 
    = 
    n_H(0)
    \sigma_\mathrm{T}
    \int \df z~
    x_e(z)
     \frac{(1+z)^2}{H(z)},
\end{align}
where $n_H(0)$ is the present number density of the hydrogen and $\sigma_\mathrm{T}$ is the Thomson scattering cross section.
At low redshifts, the number density of electrons decreases, and hence $\tau$ is less dependent on $x_e(z)$.
Thus, the optical depth is sensitive to the modification of $x_e(z)$ at high redshifts.
The Thomson scatterings decrease temperature fluctuations and increase  $E$-mode polarization. 
We calculate the modified CMB power spectrum by using the \textsf{CAMB}~\cite{Lewis:1999bs} code with the modified \textsf{RECFAST} code.
The Fig.~\ref{fig_delCl} show the modification of the CMB power spectrum by the energy  injection, and the vertical axes represent
\begin{align}
    \Delta_{C_l}  = \frac{C_{l}-C^{(0)}_{l}}{C^{(0)}_{l}}
    ,
    \label{eq_diff_CTT}
\end{align}
where $C_{l}$ and $C^{(0)}_{l}$ are the CMB power spectra with and without the energy injection. 
We plot the temperature mode (top panel) and the $E$-mode polarization (bottom panel) modified by the annihilation of the dark matter with $m_\chi=1~\mathrm{TeV}$.
The black solid line represents the modified power spectrum by the inhomogeneous dark matter with $\braket{\sigma v} =  3\times 10^{-26}\mathrm{cm}^3/\mathrm{s}$ and the same boost factor as the red solid line in Fig.~\ref{fig_compare_boostfactor}.
For comparison, we also plot the modified power spectrum by the homogeneous dark matter with $\braket{\sigma v} =  1\times 10^{-24}\mathrm{cm}^3/\mathrm{s}$ (black dotted line).
Although we use the $\mathcal O(10^{-2})$ times smaller cross section for the inhomogeneous case than the homogeneous case, both cases have comparable results because of the boost factor  Eq.~\eqref{eq_boost_factor}.
The power spectra for both cases have different scale dependence since the redshift dependence of the energy injection is different for the homogeneous and inhomogeneous dark matter, where the latter injects more energy in a later time as shown in Eq.~\eqref{fig_compare_boostfactor}.

Next, we investigate when the inhomogeneous energy absorption mainly contributes to the CMB power spectrum.
Note that the boost factor has an uncertainty in the low redshift since we neglect the effect of the standard structure formation.
We evaluate the power spectrum by artificially limiting the energy consumption epoch: early time ($z>100$, red line), middle time ($100>z>10$, green line), and late time ($10>z$, blue line).
We find that the early time contribution dominates the modification of the power spectrum for both the temperature and the polarization modes in the case of inhomogeneous energy injection, which is consistent with the fact that the CMB power spectrum is most sensitive to the energy injection at $z\sim 600$~\cite{finkbeiner2012searching}.
Thus we conclude that the CMB power spectrum puts reliable constraints on the energy injection by the inhomogeneous dark matter independent of the late time cosmology.

\section{Constraints on small-scale density fluctuation}
\label{sec_discussion}

When the dark matter abundance is determined by the thermal freeze out scenario with the $s$-wave annihilation, the cross section is estimated as $\braket{\sigma v} = 3\times 10^{-26}~\mathrm{cm}^3/\mathrm{s}$.
The current upper bound on the dark matter annihilation cross section from the CMB observation is $\braket{\sigma v}/m_\chi \lesssim  10^{-27}\mathrm{cm}^3/\mathrm{s}/\mathrm{GeV}$ for the $\chi\chi\to e^+e^-$ channel annihilation~\cite{Slatyer:2015jla} assuming the spatially homogeneous distribution of dark matter. Therefore, the thermal freezeout scenario is consistent with the CMB observation for the dark matter mass larger than $\sim 30$\,GeV.
If the small-scale curvature perturbation is large enough, on the other hand, minihalos are formed and the annihilation signal may be significantly enhanced.
Thus we can put the upper bound on the small-scale curvature perturbation by the CMB observation.
In the following, we take $m_\chi=1~\mathrm{TeV}$ and $\braket{\sigma v} = 3\times 10^{-26}\mathrm{cm}^3/\mathrm{s}$ as reference values.

We perform the Markov Chain Monte Carlo method (MCMC) with the \textsf{CosmoMC} code~\cite{Lewis:2002ah} with the modified \textsf{CAMB} code to estimate the upper limit on small-scale curvature fluctuations.
We assume the base-$\Lambda$CDM model including neutrinos with normal mass hierarchy and $\sum m_\nu = 0.06$\,eV used in the Planck collaboration~\cite{Planck:2018vyg}, and the energy injection by the dark matter annihilation is added to the model.

In the MCMC, the eight cosmological parameters are varied $\{\Omega_b h^2,~\Omega_c h^2,~100\theta_{\rm MC},~\tau,~ n_s$, ${\rm{ln}}(10^{10} A_s)$, $\log_{10}(k_*)$, $\log_{10}(A^2)\}$, where the latter two parameters determine the power spectrum of the small-scale curvature fluctuations~ (\ref{eq_powerspec_formula}).
We treat the energy injection as a perturbation on the standard $\Lambda$CDM model, and the energy consumption ratio on the thermal plasma and the boost factor are calculated based on the Planck best-fit values.
We use the top-hat prior on $\log_{10}(k_*/\mathrm{Mpc}^{-1})$ and $\log_{10}(A^2)$, and choose an appropriate parameter range for each parameter. 
We use the following data sets: the Planck 2018 data set including low-l ($l\leq 29$) TT- and EE-mode, high-l ($l\geq 30$)  TT-, TE- and EE-mode, and the CMB lensing power spectrum~\cite{Aghanim:2019ame}, and the other cosmological data on galaxy correlation functions including the baryon acoustic oscillation (BAO)
\cite{York:2000gk,Anderson:2013zyy,Ross:2014qpa,beutler20116df} and Dark Energy Survey (DES) ~\cite{Abbott:2017wau}.

\begin{figure}[th]
	\centering
	\includegraphics[width=0.45\textwidth ]{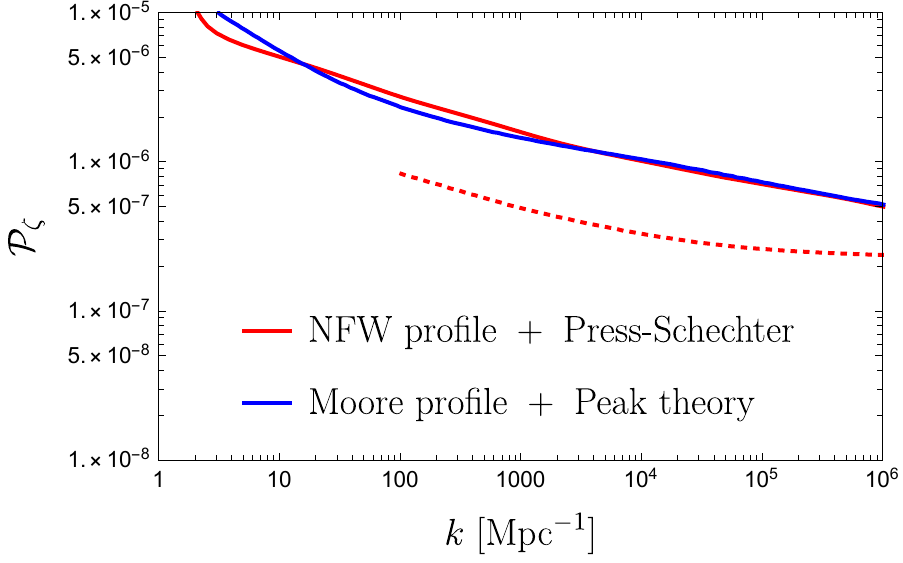}
	\includegraphics[width=0.45\textwidth ]{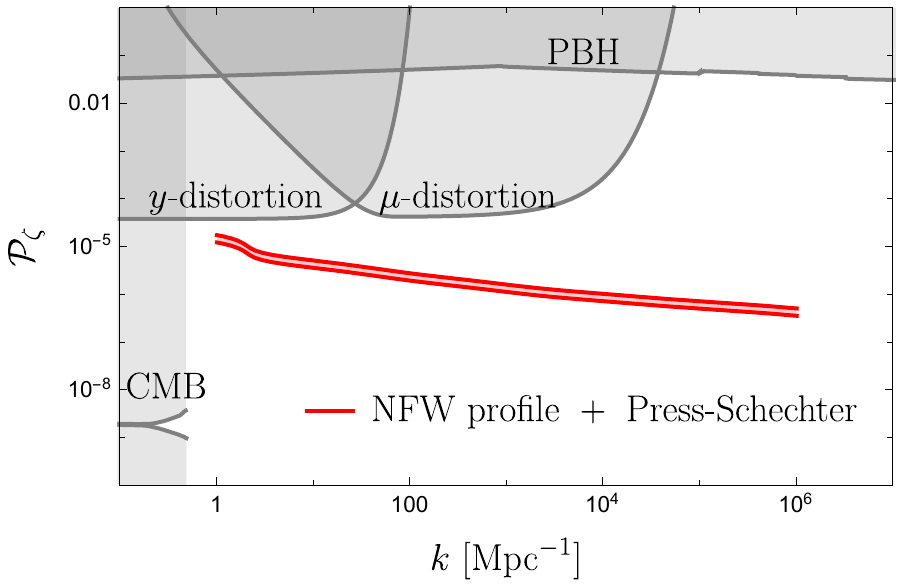}
	\caption{
    The constraints on the small-scale curvature fluctuation from the CMB observation.
    We assume the dark matter with $\braket{\sigma v} = 3\times 10^{-26}\mathrm{cm}^3/\mathrm{s}$ and $m_\chi=1~\mathrm{TeV}$ for solid lines, and $m_\chi=100~\mathrm{GeV}$ for a dotted line.
    The red and blue lines represent the constraints by the halo formation estimated for the NFW profile with the Press-Schechter formalism as discussed in Sec.~\ref{sec_halo_NFW} and the Moore profile with the peak theory as discussed in Sec.~\ref{sec_halo_Moore}.
    In the right panel, we also show other constraints from the abundance of primordial black holes~\cite{Josan:2009qn}, CMB spectral distortions~\cite{Chluba:2012we} by the COBE/FIRAS~\cite{Fixsen:1996nj}, and the CMB power spectrum by the Planck satellite~\cite{Planck:2018jri}.
	}
	\label{fig_const_power_spec}
\end{figure}

We show the results of the MCMC analysis in Fig.~\ref{fig_const_power_spec}, where each line represents the 95\% exclusion contours.
The halo formation is estimated by the methods of the NFW profile with the Press-Schechter formalism (red solid line) and the Moore profile with the peak theory (blue line).
Both methods predict a similar upper limit on the small-scale curvature fluctuation as we confirmed in Fig.~\ref{fig_compare_boostfactor}.
Our upper bound is comparable to or slightly weaker than that by the extragalactic gamma rays~\cite{Nakama:2017qac,Delos:2018ueo}.

The red dotted line represents the exclusion contour for the dark matter with mass $m_\chi=100~\mathrm{GeV}$.
Since the annihilation signal is inversely proportional to the dark matter mass, 
the constraint for $m_\chi=100~\mathrm{GeV}$ is stronger than that of $m_\chi=1~\mathrm{TeV}$.
However, the constraint on the power spectrum is less dependent on the property of minihalos and the mass of the dark matter since the halo abundance exponentially depends on the power spectrum.
The boost factor for $A^2\gtrsim 10^{-6}$ increases steeply at $z\sim 500$, and it mainly determines the upper bound on the small-scale density perturbation.
The constraint becomes weaker for smaller $k$ as we confirmed in Fig.~\ref{fig_compare_boostfactor}.
The wiggling behavior for a small $k$ may be statistical fluctuations of the MCMC analysis while the power spectrum on $\mathcal O(\mathrm{Mpc}^{-1})$ are severely constrained by observations of large-scale structure.

We comment on the baryonic effect on the constraint.
Our estimation of minihalos neglects the baryonic pressure in the halo formation, which may lead to an overestimation of density fluctuations by a factor $\Omega_{m0}/\Omega_{c0}$.
Thus, our constraint has uncertainty on the power spectrum by $(\Omega_{c0}/\Omega_{m0})^2$, which is showed by a red band in the right panel of Fig.~\ref{fig_const_power_spec},
In the right panel of Fig.~\ref{fig_const_power_spec}, we also show other constraints on the curvature power spectrum from the CMB power spectrum by Planck~\cite{Planck:2018jri}, CMB energy distortions~\cite{Chluba:2012we} by the COBE/FIRAS~\cite{Fixsen:1996nj}, and the abundance of primordial black halos~\cite{Josan:2009qn}.

\begin{figure}[th]
	\centering
	\includegraphics[width=0.4\textwidth ]{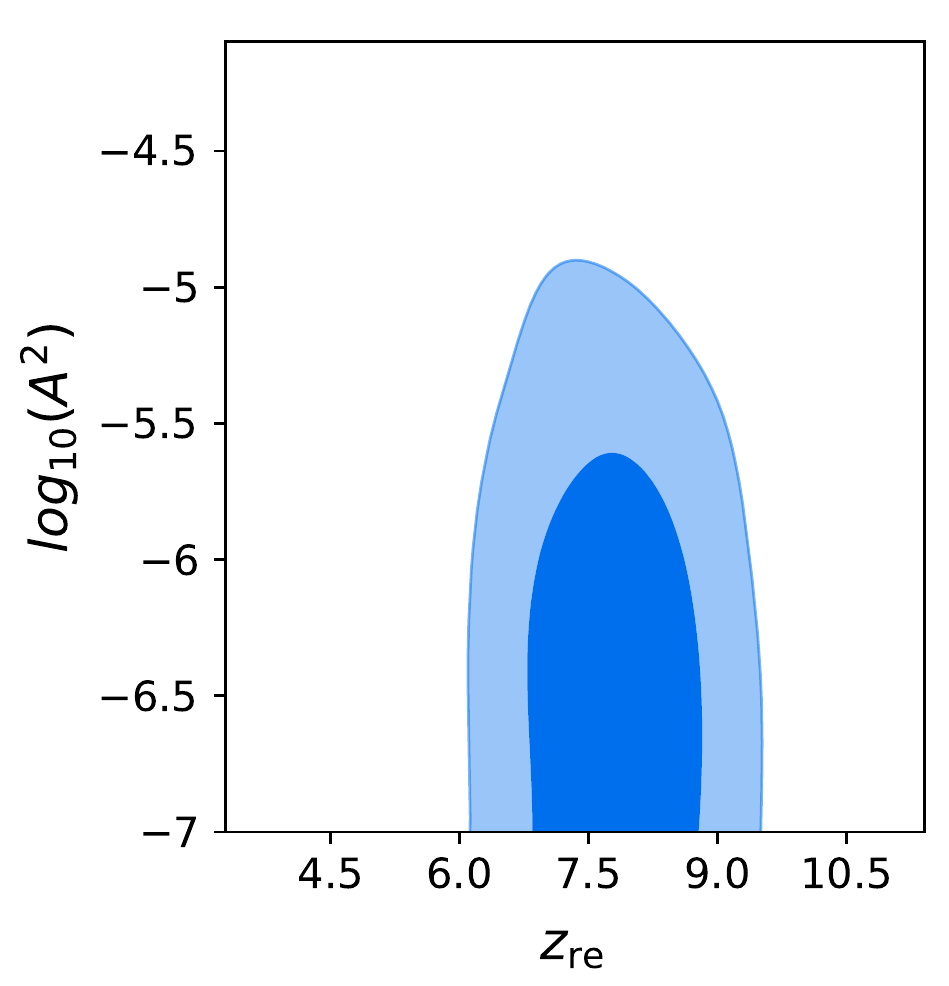}
	\caption{
    The two-dimensional contour plot of the redshift of reionization and the amplitude of the small-scale power spectrum.
    The two contours represent the $68\%$ and $95\%$ exclusion limits.
    We use the same parameter to the red solid line of Fig.~\ref{fig_const_power_spec}. 
	}
	\label{fig_2d_zre_A2}
\end{figure}

We check the degeneracy between the energy injection and the reionization.
Since both effects increase the optical depth, they may degenerate in the MCMC analysis.
The Fig.~\ref{fig_2d_zre_A2} shows the two-dimensional contour plot on the redshift of reionization, $z_\mathrm{re}$, and the amplitude of the power spectrum, $A^2$, from the MCMC analysis for the NFW profile with the Press-Schechter formalism (red line of the Fig.~\ref{fig_const_power_spec}).
$z_\mathrm{re}$ is derived by other cosmological parameters assuming the \textsf{tanh} model of reionization.
The degeneracy of the parameters is solved in Fig.~\ref{fig_2d_zre_A2} since the exotic energy injection and the reionization effect have different redshift dependence.
A similar result is confirmed in the case of the exotic energy injection by the mixed WIMP and the primordial black hole scenarios ~\cite{Tashiro:2021xnj}.

\section{Conclusion}
\label{sec_conclusion}

Although the CMB observations have revealed the primordial fluctuations on scales larger than $\mathcal O(1)\mathrm{Mpc}$, small-scale fluctuations are still less constrained.
When the small-scale density fluctuation is much larger than the naive extrapolation from the CMB scales with the assumption of scale-invariant power spectrum, minihalos are formed in the early stage of the universe.
The minihalos amplify the inhomogeneity of the dark matter distribution, which increases the annihilation signal of the dark matter.
In the standard thermal freeze out scenario of dark matter production, the cross section is expected to be  $\braket{\sigma v} =  3\times 10^{-26}\mathrm{cm}^3/\mathrm{s}$ for the $s$-wave annihilation.
The dark matter annihilation injects high-energy particles into the thermal plasma, and they act as extra ionization sources of the neutral hydrogen and helium.
The modified ionization history is constrained by the observation of the CMB anisotropy.
Such annihilation signals from minihalos are intensively studied in the context of the extragalactic gamma rays~\cite{Nakama:2017qac,Delos:2018ueo}, while there are fewer works that focused on a constraint from the CMB anisotropy.

In this work, we have investigated the constraints on small-scale curvature fluctuations from the CMB anisotropy through the dark matter annihilation.
We estimated the abundance and the profile of minihalos taking into account the recent numerical simulation~\cite{Delos:2018ueo}.
We have adopted two different methods to estimate the properties of the minihalos.
First, the halo abundance was estimated by the  Press-Schechter formalism assuming the NFW profile~\cite{Nakama:2017qac}, which is similar to the method often used for the standard halos.
Second, following the numerical simulation~\cite{Delos:2018ueo} we took the picture that minihalos have the Moore profile and their formation time determines their property~\cite{Delos:2018ueo}. 
It was found that both models highly amplify dark matter annihilation rate as shown in Fig.~\ref{fig_compare_boostfactor}.
We estimated the energy injection from the dark matter annihilation assuming the standard thermal freezeout scenario.
The modified ionization history was calculated by the \textsf{RECFAST} code, where we precisely included the fraction of the injected energy used for ionization taking into account all relevant scattering processes.
Using the modified \textsf{RECFAST} code, we calculated the CMB power spectrum by the \textsf{CAMB} code and performed the MCMC analysis by the \textsf{CosmoMC} code to put the exclusion limit of the parameters.
The resultant constraints on the small scale power spectrum of the curvature fluctuation are shown in Fig.~\ref{fig_const_power_spec} and comparable to that from the extragalactic gamma rays~\cite{Nakama:2017qac,Delos:2018ueo}.
We have confirmed that our constraints mainly come from the ionization process at the high redshifts, $z>100$.
Therefore, our constraint is independent of some uncertainty of the halo properties in the late times, e.g. the destruction of the minihalos by the standard halos~\cite{Nakama:2017qac}.
Thus, the CMB constraint on the small scale curvature fluctuations derived in this paper is reliable and complementary to the constraint from the extragalactic gamma rays.

\section*{Acknowledgments}

This work was supported by JSPS KAKENHI Grant Nos. 17H01131 (M.K.), 17K05434 (M.K.), 20H05851(M.K.), 21K03567(M.K.), JP19J21974 (H.N.), 18K03609 (K.N.), 17H06359 (K.N.), 
Advanced Leading Graduate Course for Photon Science (H.N.), and 
World Premier International Research Center
Initiative (WPI Initiative), MEXT, Japan (M.K.).

\appendix

\small
\bibliographystyle{apsrev4-1}
\bibliography{Ref}

\end{document}